\documentclass[reprint,aps,prl,twocolumn,superscriptaddress,showkeys,showpacs,longbibliography]{revtex4-1}
\usepackage{graphicx}
\usepackage{amssymb}
\usepackage{float}
\usepackage[T1]{fontenc}
\usepackage[utf8]{inputenc}


\graphicspath {{./figs/}}
\makeatletter
\def\input@path{{./figs/}}
\makeatother

\begin{document}
\title{Accelerating molecular dynamics simulations with population annealing}
\author{Henrik Christiansen}
\email{henrik.christiansen@itp.uni-leipzig.de}
\affiliation{Institut für Theoretische Physik, Universität Leipzig, Postfach
  100\,920, 04009 Leipzig, Germany}
\author{Martin Weigel}
\email{martin.weigel@complexity-coventry.org}
\affiliation{Applied Mathematics Research Centre, Coventry University, Coventry 
CV1 5FB, England}
\author{Wolfhard Janke}
\email{wolfhard.janke@itp.uni-leipzig.de}
\affiliation{Institut für Theoretische Physik, Universität Leipzig, Postfach
  100\,920, 04009 Leipzig, Germany}
\date{\today}

\begin{abstract}
  Population annealing is a powerful tool for large-scale Monte Carlo simulations. We
  adapt this method to molecular dynamics simulations and demonstrate its excellent
  accelerating effect by simulating the folding of a short peptide commonly used to
  gauge the performance of algorithms. The method is compared to the well established
  parallel tempering approach and is found to yield similar performance for the same
  computational resources. In contrast to other methods, however, population
  annealing scales to a nearly arbitrary number of parallel processors and it is thus
  a unique tool that enables molecular dynamics to tap into the massively parallel
  computing power available in supercomputers that is so much needed for a range of
  difficult computational problems.
\end{abstract}

\maketitle

Simulations of complex systems with rugged free-energy landscapes are among the
computationally most challenging problems \cite{janke2007rugged}. Next to structural
and spin glasses, macromolecules including proteins are prototypical examples of
systems were frustration results in many (free) energy minima separated by
barriers. A range of methods has been developed to overcome the problem of the system
getting trapped in a local minimum \cite{berg:92b,hukushima:96a,wang:01a,laio:02}.
The most popular choice is parallel tempering
\cite{swendsen:86,geyer:91,hukushima:96a} (also known as replica exchange) which has
been shown to successfully sample a broad configuration space when applied to
peptides \cite{hansmann:97,sugita:99}. This method uses a small number of replicas
which are, a priori, simulated independently at different temperatures.  At regular
intervals the replicas exchange configurations with a probability adjusted to their
relative Boltzmann weight.  Although this approach is easily parallelized, the number
of processors that can be reasonably used is limited by the increasing time it takes
for a system to traverse the whole temperature range when the number of temperature
points increases.

Population annealing
\cite{iba:01,hukushima:03,machta:10a,wang:15a,barash:17,callaham:17} is another
generalized-ensemble simulation scheme, which was originally introduced for Monte
Carlo simulations. While it was found to be similarly good at dealing with complex
free-energy landscapes, it can easily make use of many thousands of processors
including GPUs \cite{barash:16} and scales extremely well. The approach is based on the
ideas of sequential Monte Carlo methods. It consists of setting up an ensemble of $R$
independent configurations at a high temperature where equilibration is
straightforward. The population is then sequentially cooled in small steps. At each
temperature, the population is resampled according to a ratio of Boltzmann weights
and then evolved for a number of simulation steps. This keeps the population in
equilibrium and thus observables can be calculated as population averages at each
temperature. Parallelization is over independent replicas, and this allows for the
good scaling properties as population sizes $R$ as large as $10^6$ and beyond are not
uncommon. Population annealing has been shown to perform well, e.g., in Monte Carlo
simulations of spin glasses \cite{hukushima:03,wang:15a}. Here we show how it can be
adapted to molecular dynamics simulations and thereby unleash the power of massively
parallel computations for a wide range of applications, for instance in the simulation of
(biological) macromolecules, structural glasses, or colloidal systems.

The population annealing (PA) method proposed here is a straightforward extension of
canonical (typically NVT) molecular dynamics (MD) simulations. It achieves
substantially improved equilibration properties by following a population of systems
that are independently evolved with the underlying MD algorithm while periodically
replicating particularly well equilibrated copies (reminiscent of ``go with the
winners'' strategies \cite{grassberger:02a}). The temperature annealing run starts
with a population of $R$ replicas that are initially equilibrated at some high
temperature $T_0$ where relaxation times are small.  The actual annealing process is
then implemented by successively lowering the temperature from $T_{i-1}$ to
$T_i < T_{i-1}$, resampling the population with the relative Boltzmann weight, and
simulating each new replica independently for $\theta$ MD steps.

This scheme for population annealing molecular dynamics (PAMD) simulations can be
summarized as follows:
\begin{enumerate}
\item Set up an equilibrium ensemble of $R$ independent copies of the system at some
  high temperature $T_0$.
\item Resample the ensemble of systems to a temperature $T_i<T_{i-1}$ by replicating
  each copy a number of times proportional to the relative Boltzmann weight
  $\tau_j = e^{-(1/k_BT_i-1/k_BT_{i-1})E_j}/Q$, where
  $Q=\sum_{j=1}^R e^{-(1/k_BT_i-1/k_BT_{i-1})E_j}/R$ is a normalization factor. The
  velocities are taken care of through rescaling by a factor $\sqrt{T_i/T_{i-1}}$,
  such that what enters the resampling probability is only the potential energy
  $E_j$.
\item Update each copy with $\theta$ simulation steps of the underlying MD 
algorithm.
\item Calculate observables $\mathcal{O}$ at temperature $T_i$ as population averages.
\item Goto step 2 until $T_i$ reaches or falls below the target temperature $T_N$.
\end{enumerate}
Note that the replication in step 2 includes the case of making zero copies,
corresponding to pruning the corresponding configuration from the population. While a
number of different implementations for the resampling step are possible
\cite{machta:10a,wang:15a,barash:16}, we here draw $R$ samples from a multinomial
distribution with probabilities $\tau_j$, $j=1,\ldots,R$. This ensures that the total
population size is constant, thus simplifying parallel implementations on distributed
machines. The choice of temperatures $T_i$ follows the same rules as for the
parallel tempering (PT) method, i.e., a sufficient overlap of the energy histograms is
required. The temperatures can also be chosen automatically based on an overlap
condition \cite{barash:16,weigel:17a,christiansen:19}.

\begin{figure}[!tb]
  \includegraphics{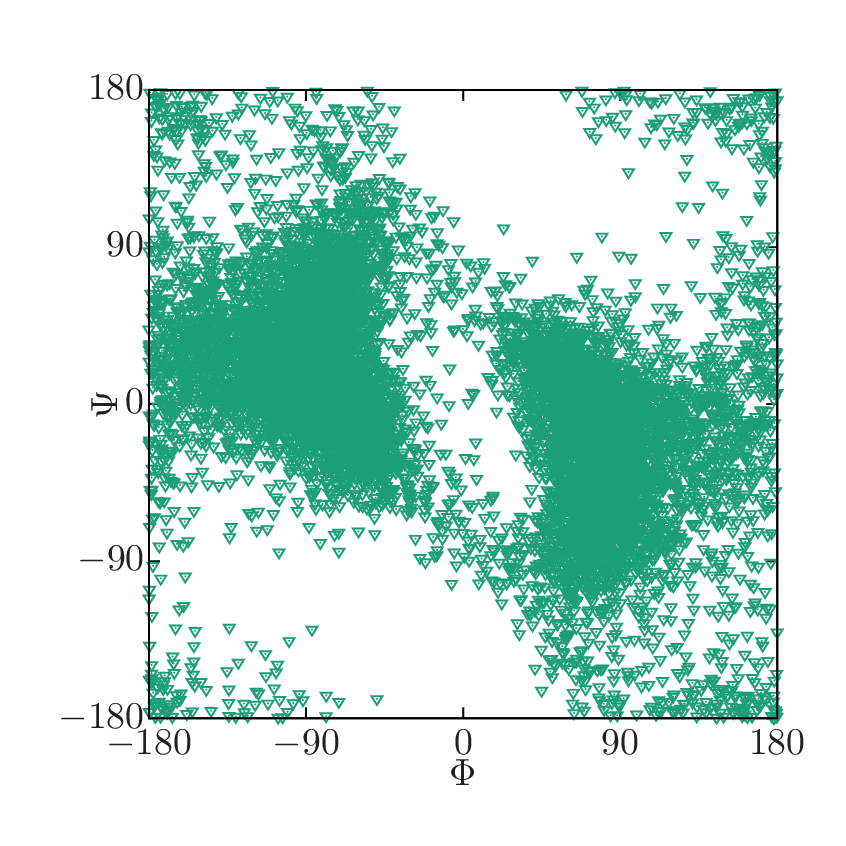}
  \caption{Ramachandran plot for the amino acid GLY-2 of met-enkephalin at 
    $T_0=700$~K obtained from a canonical simulation.}
  \label{fig:rama-high}
\end{figure}

The vast majority of simulation time is spent on MD steps advancing the
replicas. To get good performance there, we rely on the OpenMM package
\cite{eastman:17} which can also make use of GPUs. It is important to note that in
PAMD simulations the thermostat has to be stochastic in nature: If more than one copy
of a replica is created during resampling these are identical initially and the noise
from the thermostat is required to make sure that they decorrelate over time.  Noise
sources that lead to a stronger perturbation of the microcanonical trajectories are
expected to increase the overall efficiency of the method due to a stronger decorrelation.
This effect can be varied, e.g., by tuning the stochastic collision
frequency in the Andersen thermostat \cite{christiansen:18b}. Additionally one can 
make use of schemes such as the Lowe-Andersen thermostat \cite{lowe:99,majumder:18},
which cannot be efficiently implemented in combination with domain decomposition.

The possibility of creating more than one copy implies that, after resampling, there
are correlations between members of the population. These are relevant for the
analysis of statistical errors and the level of exploration of configuration space.
Their influence can be assessed via binning over the population, for details see
Ref.~\citenum{weigel:17a}.

During the PAMD run the values of the normalizing factors $Q$ are separately
recorded. As is easily seen \cite{machta:10a,barash:16}, these estimate the (configurational)
partition function ratios $ Q(T,T') \approx \mathcal{Z}(T)/\mathcal{Z}(T')$ or,
equivalently, the free-energy differences
$\mathcal{F}(T')/k_B T' - \mathcal{F}(T)/k_B T = \ln \mathcal{Z}(T) - \ln \mathcal{Z}(T')$. This allows
us to not only determine free energies directly, but also to use a straightforward implementation of the weighted
histogram analysis method (WHAM) \cite{ferrenberg:89a,kumar:92}. In practice, we
refine the results by a few iterations of the self-consistency equations of
Refs.~\citenum{ferrenberg:89a,kumar:92}, but the above procedure already provides
excellent starting values.

\begin{figure}[!tb]
  \includegraphics{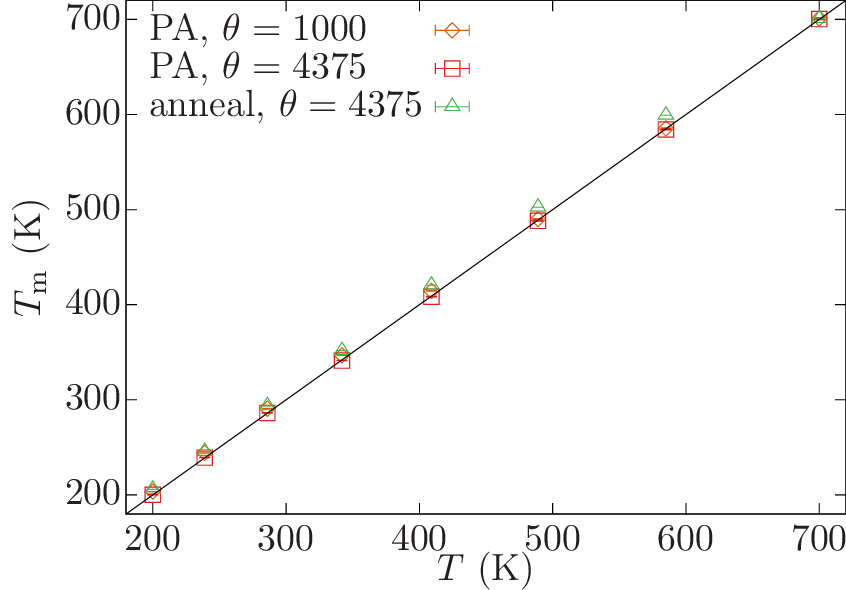}
  \caption{Measured temperature $T_\mathrm{m}$ vs.\ the heat-bath temperature 
    $T$ for population annealing (PA) simulations with $\theta = 1000$ and $\theta 
    = 4375$ as well as for PA with resampling turned off (``anneal'').}
  \label{fig:temp}
\end{figure}

As a test system for demonstrating the efficiency of our method we consider the
penta-peptide met-enkephalin in vacuo.  For this peptide we cap the ends with acetyl
respectively N-methyl groups. Met-enkephalin has the amino-acid sequence
Tyr-Gly-Gly-Phe-Met. Below we will focus on the quality of the relaxation of the
dihedral angles in the inner amino acids GLY-2, GLY-3, and PHE-4.  To model the
interactions we employ the AMBER force-field ff94 \cite{cornell:95}. The temperature
set is adapted from a PT simulation of the same peptide \cite{sugita:99} and reads
$T_i=700$, $585$, $489$, $409$, $342$, $286$, $239$, $200$~K.  We initialize the
population of replicas from a canonical, well-equilibrated simulation at
$T_0 = 700$~K. The distribution of dihedral angles $\Phi$ and $\Psi$ of GLY-2 from
this initial simulation is shown in the Ramachandran plot of
Fig.~\ref{fig:rama-high}. A full PT simulation results in a virtually identical angle
distribution, clearly proving equilibration at $T_0$.

For the PAMD simulation we take configurations from this initial run at $T_0$ to
initialize $R=10\,000$ replicas, and additionally apply $\theta$ MD steps each to
ensure statistical independence (the choice of $\theta$ is discussed below).  Unlike
for the Monte Carlo variant of PA, here $T_0$ is finite.  We used the stochastic
Langevin thermostat with a different random-number seed for each replica. The
friction coefficient is $\gamma=1/$ps and the MD algorithm is a variant of velocity
Verlet with time step $dt=0.5$~fs. Figure~\ref{fig:temp} shows the measured vs.\ the
given heat-bath temperature for $\theta = 1000$ and $\theta = 4375$, respectively, as
well as for an annealing scheme with $\theta = 4375$ without resampling. A closer
look at this plot reveals that $\theta = 1000$ MD steps are not sufficient for
equilibration but $\theta = 4375$ are, such that we use $\theta = 4375$
subsequently. The plot also shows that omitting the resampling step of PAMD,
corresponding to a pure annealing of the population of independent runs, worsens
the equilibration behavior.

\begin{figure}[!tb]
  \includegraphics{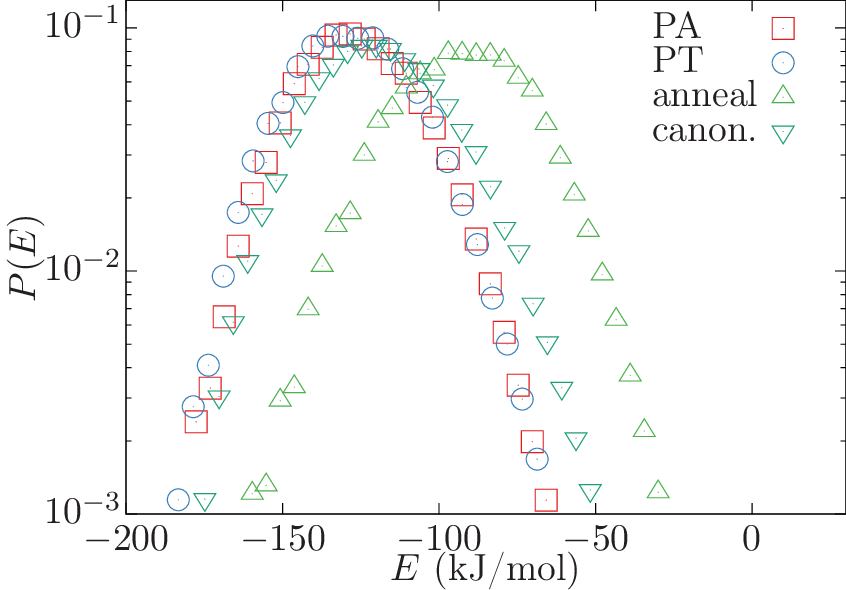}
  \caption{Energy histograms at the lowest temperature, $T=200$~K, as obtained from
    the population annealing (PA), parallel tempering (PT), PA without resampling
    (``anneal''), and canonical simulations, respectively.}
  \label{fig:hist}
\end{figure}

For the actual benchmark, we compared PAMD to a sequence of canonical simulations, to
PT, and to PAMD without resampling, all with the same temperature set. To ensure
fair rules for the competition, all methods were assigned the same computational
budget corresponding to a total of $200$~ns of MD simulation --- the computational
overhead of the swaps in PT and the resampling steps in PA is negligible.  For each
simulation method, the budget is distributed differently.  In the canonical
simulations the protein was equilibrated for $12.5$~ns for each of the $8$
temperatures, followed by another $12.5$~ns of simulation during which $10\,000$
measurements were taken.  For PA we used the configurations from this canonical
simulation at $T_0$ as our $R=10\,000$ initial configurations (see
Fig.~\ref{fig:rama-high}), and subjected them to an additional $\theta = 4375$ MD
steps each to improve decorrelation.  The actual simulations (including the initial
$4375$ MD steps) then ran for a total of $175$~ns, corresponding to the above
mentioned $4375$ MD steps per replica and temperature.  In PT a total of $50$~ns were
spent for equilibration.  The remaining $150$~ns were equally distributed on the $8$
temperatures. Measurements were taken before each PT exchange. In total, this step
was performed $10\,000$ times, amounting to $3750$ MD steps between exchanges.

\begin{figure}[!bt]
  \includegraphics{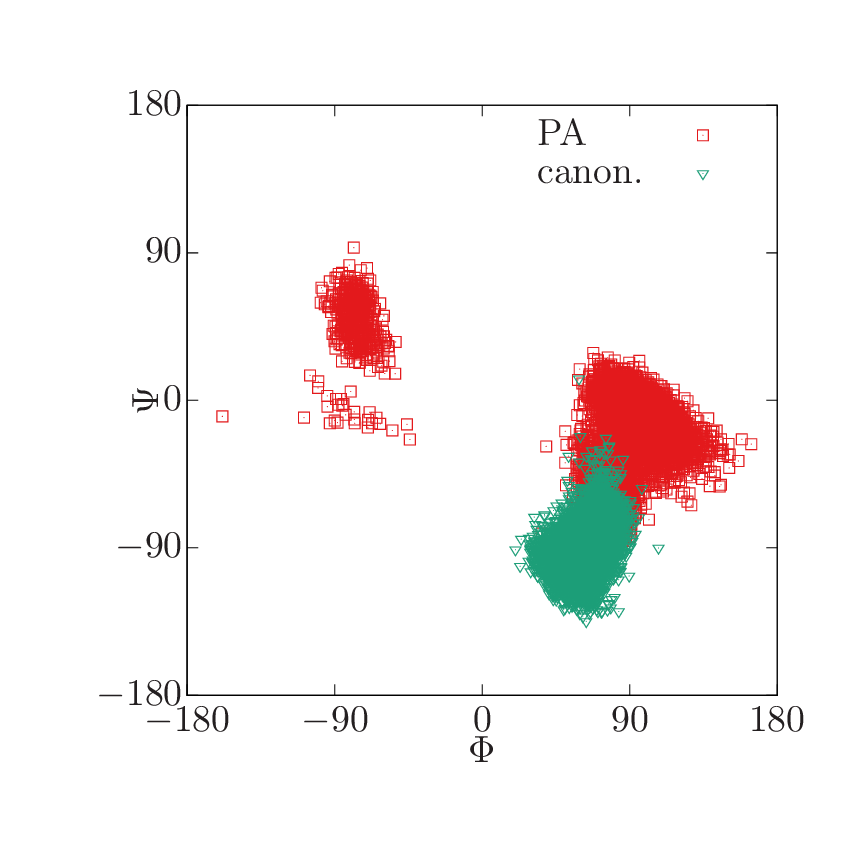}
\caption{Ramachandran plot for GLY-3 at $T=200$~K obtained from the population annealing 
  (PA) and a canonical simulation. The total run time of both methods was $200$~ns. The 
  configuration space sampled by PA is far superior to that sampled by a canonical simulation.}
\label{fig:rama-low-can}
\end{figure}

\begin{figure*}[!bt]
    \includegraphics{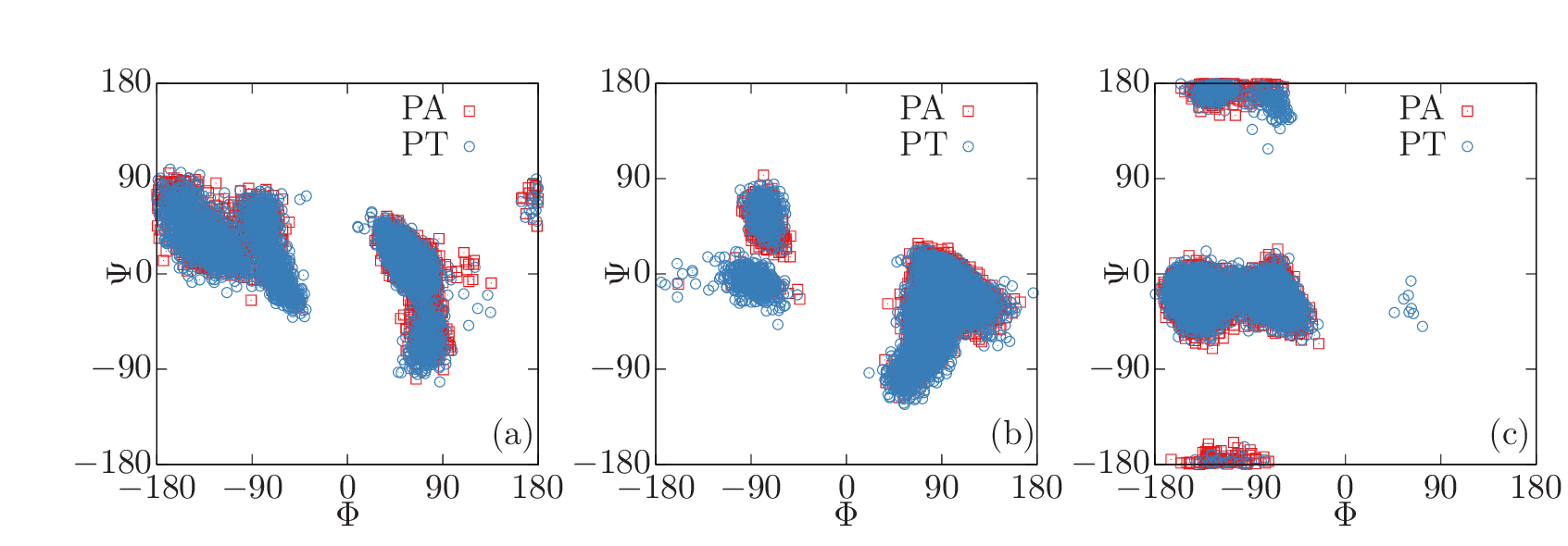}
  \caption{Ramachandran plots for (a) GLY-2, (b) GLY-3, and (c) PHE-4 at 
    $T=200$~K obtained from the population annealing (PA) and parallel tempering (PT)
    simulations. The total amount of parallel work for each method was 
    the equivalent of $200$~ns of MD steps. The fraction of configuration space 
    sampled by the two methods is approximately equivalent.
  }
  \label{fig:rama-pt}
\end{figure*}

In Fig.~\ref{fig:hist} we present the distribution of potential energy for
met-enkephalin at the lowest temperature $T=200$~K from PA with and without resampling, 
PT as well as canonical simulations. The histograms obtained from PA and PT
are compatible with each other, while both the canonical simulation and PA without
resampling are shifted towards higher energies.  For the canonical simulation this
indicates trapping in a local minimum. The annealing simulation distribution
indicates that without resampling the $\theta=4375$ MD steps are not sufficient to
keep the system in equilibrium at the given cooling rate, consistent with the
observation in Fig.~\ref{fig:temp}.

Figure~\ref{fig:rama-low-can} shows a Ramachandran plot for GLY-3 at the lowest
temperature $T=200$~K from PAMD as compared to the canonical simulation.  As
expected, the distributions of the dihedral angles $\Phi$ and $\Psi$ for the two sets
of simulations differ substantially. While the canonical simulation got trapped and
thus only simulated a fraction of the available conformations, PAMD sampled a wide
configuration space. With the same amount of computational resources thus a much
better sampling is achieved.

While PAMD is hence clearly superior to a purely canonical simulation, it is
important to check how it competes against PT as the {\em de facto\/} standard for
accelerated simulations. This comparison can be found in Fig.~\ref{fig:rama-pt},
again showing Ramachandran plots of the three inner amino acids GLY-2, GLY-3, and PHE-4 at
$T=200$~K. It is evident that both methods sample a wide configuration space with a
few areas having been discovered only by PT and others only by PA. Overall, the
quality of sampling for this system is comparable between PT and PA.

\begin{figure}[!tb]
  \includegraphics{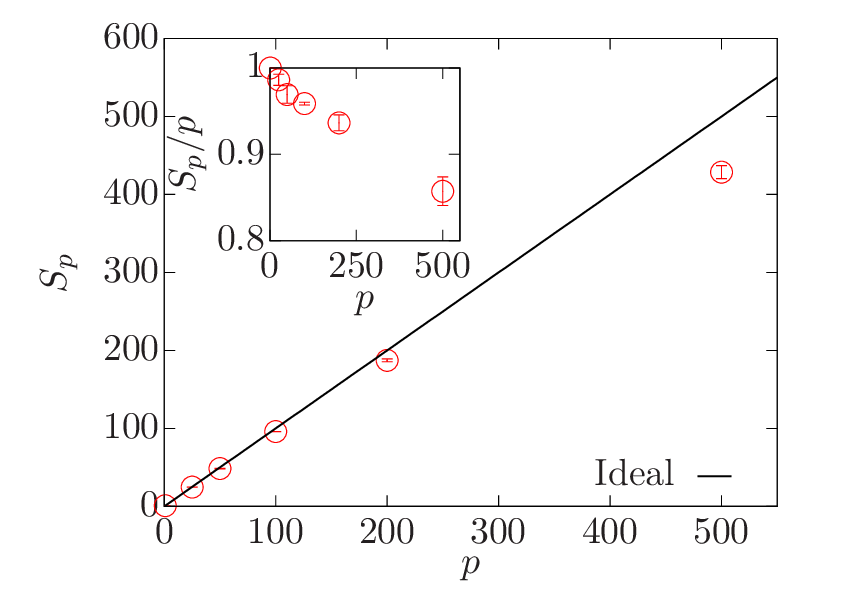}
  \caption{
    Speedup $S_p$ for a different number of processors $p$ for the PAMD with $\theta=4375$
    and $R=10000$. The solid line corresponds to perfect scaling. In the inset, we show the
    efficiency $S_p/p$ for the same simulation.
  }
  \label{fig:scaling}
\end{figure}

The crucial advantage of PAMD lies in the possibility to improve it essentially
arbitrarily by using parallel resources. To illustrate this, we show in
Fig.~\ref{fig:scaling} the speedup $S_p$ when increasing the number of cores $p$. As
the inset illustrates, the scaling efficiency $S_p/p$ is consistently high and never
drops below 85\% in the range considered here. We note that for a ``strong scaling''
scenario where the population size is increased in line with the computational
resources and one hence studies the same system but with decreasing biases and
increasing statistical accuracy \cite{weigel:17a}, the scaling properties are even
better than for the ``weak scaling'' demonstrated in Fig.~\ref{fig:scaling}. A similar
improvement cannot be achieved from additional parallel resources in PT as combining
the results of several independent PT runs does not improve thermalization. 

The possibility of PA to faithfully sample systems with many
different minima requires a sufficiently large population to occupy all of the
relevant valleys in the free-energy landscape. This is similar to the requirement in
PT of sufficiently long runs to achieve fair sampling, but while larger populations
can be treated in parallel, longer runs cannot. In that sense time in PT corresponds
to population size in PA.

\begin{figure}[!tb]
  \includegraphics{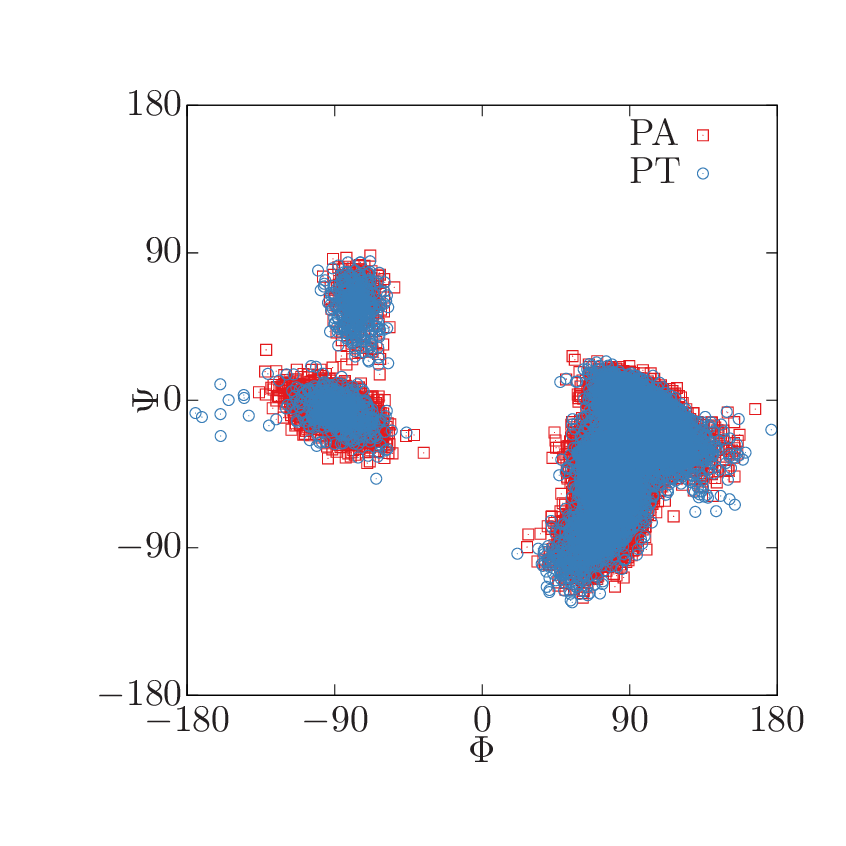}
  \caption{ Ramachandran plot for GLY-3 at $T=200$~K obtained from a population
    annealing (PA) simulation with $R=2 \times 10^4$ replicas as compared to the
    parallel tempering (PT) simulation of total run time $200$~ns already shown in
    Fig.~\ref{fig:rama-pt}.  }
  \label{fig:rama-long}
\end{figure}

The improvement in the quality of results with population size is illustrated in
Fig.~\ref{fig:rama-long}, where we compare the results for GLY-3 of a PAMD simulation
with twice the number of replicas $R=2 \times 10^4$, but otherwise using identical
simulation parameters, to the PT simulation employing $200$~ns of MD already shown in
Fig.~\ref{fig:rama-pt}.  Given sufficient parallel resources, the wall-clock time of
this enlarged PAMD simulation is approximately the same as that for the previous PAMD
run. As is clearly seen from Fig.~\ref{fig:rama-long}, however, the sampling of
configuration space is significantly improved, and the area of $\Phi \approx -90$ and
$\Psi \approx 0$ that is difficult to observe is now sampled significantly better.

In conclusion, we have shown that the combination of population annealing with
molecular dynamics simulations is a very promising new tool.  As demonstrated for the
folding of the penta-peptide met-enkephalin, it yields a broadening of the accessible
configuration space and faster relaxation on par with parallel tempering, but with
the potential of employing essentially arbitrarily large parallel resources that is
lacking in parallel tempering. For Monte Carlo simulations population annealing was already shown to
scale to millions of threads for spin systems on GPU clusters \cite{barash:16}. A
wide range of further improvements come to mind, including advanced schedules of
temperatures, sweep protocols and population sizes \cite{barash:16,weigel:17a,christiansen:19},
adaptations for other ensembles such as NVE, and combinations with umbrella
sampling. Crucially, however, already the extremely simple extension of standard
molecular dynamics simulations as presented here provides an outstanding acceleration
given sufficiently powerful parallel resources. For a broad range of systems this
opens the door to the world of highly efficient computer simulations on petaflop
supercomputers of the present and the exaflop machines of the future.

\begin{acknowledgments}
  This project was funded by the Deutsche Forschungsgemeinschaft (DFG) under Grant
  No.\ SFB/TRR 102 (project B04), and further supported by the Leipzig Graduate
  School of Natural Sciences ``BuildMoNa'', the Deutsch-Franz\"osische Hochschule
  (DFH-UFA) through the Doctoral College ``$\mathbb{L}^4$'' under Grant No.\
  CDFA-02-07, and the EU Marie Curie IRSES network DIONICOS under Grant No.\
  PIRSES-GA-2013-612707.
\end{acknowledgments}

\end{document}